\documentclass[preprint,3p,twocolumn]{elsarticle}

\usepackage[colorlinks]{hyperref}
\usepackage{tabularx}
\usepackage{graphicx}
\usepackage{amssymb,amsmath,bm,tensor,braket}
\usepackage{xcolor}
\usepackage[varg]{txfonts}
\usepackage{enumerate}
\usepackage{mathtools}
\usepackage[utf8]{inputenc}
\usepackage[normalem]{ulem}
\usepackage{lineno,hyperref}
\modulolinenumbers[5]

\journal{Physics Letters B}

\newcommand*\dd{\mathop{}\!\mathrm{d}}

\def\al{\alpha}
\def\be{\beta}
\def\ga{\gamma}
\def\de{\delta}

\def\ep{\epsilon}

\def\et{\eta}
\def\th{\theta}

\def\rh{\rho}

\def\vp{\varphi}

\def\Si{\Sigma}

\def\Om{\Omega}

\def\cL{{\cal L}}

\def\mn{{\mu\nu}}

\def\prt{\partial}

\def\sqr#1#2{{\vcenter{\vbox{\hrule height.#2pt
         \hbox{\vrule width.#2pt height#1pt \kern#1pt
         \vrule width.#2pt}
         \hrule height.#2pt}}}}

\newcommand{\beq}{\begin{equation}}
\newcommand{\eeq}{\end{equation}}
\newcommand{\bea}{\begin{eqnarray}}
\newcommand{\eea}{\end{eqnarray}}
\newcommand{\rf}[1]{(\ref{#1})}

\newcommand{\bM}{\begin{pmatrix}}
\newcommand{\eM}{\end{pmatrix}}

\begin{document}

\begin{frontmatter}

\title{Neutron Star Structure in the Minimal Gravitational Standard-Model Extension \\ and the Implication to Continuous Gravitational Waves}

\author[1]{Rui Xu\corref{cor1}}\ead{xuru@pku.edu.cn}
\author[2]{Junjie Zhao}
\author[1]{Lijing Shao\corref{cor1}}\ead{lshao@pku.edu.cn}

\cortext[cor1]{Corresponding authors}
\address[1]{Kavli Institute for Astronomy and
Astrophysics, Peking University, Beijing 100871, China}
\address[2]{School of Physics and State Key Laboratory of Nuclear Physics and Technology, Peking University, Beijing 100871, China}

\begin{abstract}
Tiny violation of Lorentz invariance has been the subject of theoretic study and
experimental test for a long time. We use the Standard-Model Extension (SME)
framework to investigate the effect of the minimal Lorentz violation on the
structure of a neutron star. A set of hydrostatic equations with modifications from Lorentz violation are derived, and then the modifications are isolated and added to the Tolman-Oppenheimer-Volkoff (TOV) equation as the leading-order Lorentz-violation corrections in relativistic systems. A perturbation solution to the leading-order modified TOV equations is found. The quadrupole moments due to the anisotropy  in the structure of neutron stars are calculated and used to estimate the quadrupole radiation of a spinning neutron star with the same deformation. The
calculation puts forward a new test for Lorentz invariance in the
strong-field regime when continuous gravitational waves are observed in the
future.
\end{abstract}


\end{frontmatter}


\section{Introduction}

Neutron stars are important objects in astrophysics as the observation of
them provides unique tests of gravitational theories and fundamental
principles~\cite{Lorimer:2005misc, Wex:2014nva, Shao:2016ezh}. The masses
and orbital parameters of highly magnetized rotating neutron stars that
emit radio waves in a binary system can be accurately determined by pulsar
timing techniques \cite{Lorimer:2005misc, Ozel:2016oaf, Lattimer:2019eez}.
Another
observation channel, the direct gravitational wave detection
from binary neutron stars, though only had its practice in 2017
\cite{TheLIGOScientific:2017qsa}, is having more facilities in construction
and developing promisingly \cite{Evans:2016mbw, LIGOScientific:2019vkc,
Michimura:2019cvl, Kuns:2019upi, Sedda:2019uro}. No
matter through pulsar signals or gravitational waves, the information of
neutron stars we receive not only supplies us knowledge about matter at
supranuclear densities \cite{Lattimer:2000nx,
Steiner:2010fz,Lattimer:2015nhk}, but also constrains various alternatives
to General Relativity (GR) \cite{Berti:2015itd, Will:1997bb, Miao:2019nhf,
Shao:2017gwu}.

We are particularly interested in testing gravitational theories with tiny
violation of Lorentz invariance, which is believed to be a possible quantum
gravity effect \cite{Kostelecky:1988zi, Kostelecky:1989px, Kostelecky:1989jp,
Kostelecky:1991ak, Kostelecky:1995qk, Collins:2004bp, Collins:2006bw,
AmelinoCamelia:2008qg}. At the low-energy level, any kind of such violation includes couplings between the Lorentz-violation fields and the
conventional fields in effective field theory \cite{Colladay:1996iz}. The
framework is called the Standard-Model Extension (SME) \cite{Colladay:1998fq,
Kostelecky:2003fs}, in which the complete Lagrangian density reads
\cite{Bailey:2014bta}
\bea
\cL_{\rm SME} = \cL_{\rm GR} + \cL_{\rm SM} + \cL_{\rm LV} + \cL_{\rm k} \, .
\label{lsme}
\eea
In the expression, $\cL_{\rm GR}$ represents the usual Einstein-Hilbert term for
General Relativity, and $\cL_{\rm SM}$ is the Lagrangian density of the Standard
Model. One of the extra terms, $\cL_{\rm LV}$, consists of Lorentz violation
couplings in the form of $\left(k^{(d)}\right)^a J_a$, with $\left(k^{(d)}\right)^a$ being a
Lorentz-violation field and $J_a$ being an operator of mass-dimension $d$
constructed from the conventional fields. The other term, $\cL_{\rm k}$, describes the
dynamics of the Lorentz-violation fields. 

The Lorentz-violation couplings in $\cL_{\rm LV}$ are naturally categorized by
the mass-dimensions of the conventional field operators. For our purpose, we
only consider the gravitational SME where the conventional field operators $J_a$
are constructed using the Riemann tensor so that $d$ starts with 4
\cite{Bailey:2014bta, Bailey:2006fd}. The $d=4$ coupling is simply
$\frac{1}{16\pi G} \left(k^{(4)} \right)^{\al\be\ga\de} R_{\al\be\ga\de} $, but intentionally
introduced in terms of the trace-free components of
$\left( k^{(4)} \right)^{\al\be\ga\de}$ and $R_{\al\be\ga\de}$ in Ref.~\cite{Bailey:2006fd}
as
\bea
\cL^{(4)}_{\rm LV} = \frac{1}{16\pi G} \left( -u R + s^\mn R^{T}_{\mn} + t^{\al\be\ga\de} C_{\al\be\ga\de} \right),
\label{mSME}
\eea
where $R^{T}_{\mn}$ is the trace-free Ricci tensor and $C_{\al\be\ga\de}$ is the
Weyl conformal tensor. By splitting the $d=4$ Lorentz-violation coefficient $\left( k^{(4)} \right)^{\al\be\ga\de}$ into three pieces, $u$, $s^{\mu\nu}$, and $t^{\alpha\beta\gamma\delta}$, Eq. \rf{mSME} consists of all the Lorentz-violation couplings in the minimal gravitational SME. Any coupling with the mass dimension $d$ of
the conventional field operator greater than 4 belongs to the nonminimal sector of the SME framework
\cite{Bailey:2014bta, Kostelecky:2015dpa, Kostelecky:2016uex,
Kostelecky:2016kfm}. Nonminimal couplings involve more derivatives and are
considered to be suppressed at least by factors of $\frac{E}{M_P}$, where $E$ is the
energy below which effective field theory works and $M_P$ is the Planck mass.
Therefore, we will only consider the minimal couplings in our work but point out
that extending the treatment to nonminimal Lorentz-violation couplings is still
a relevant topic as in some specific Lorentz-violation models there is no
minimal Lorentz-violation coupling and therefore the dominant effect comes from
nonminimal terms \cite{Carroll:2001ws, Berger:2001rm}. 

As we study gravity at the classical level, in Eq. \rf{lsme}, $\cL_{\rm SM}$ is
replaced by the Lagrangian density for macroscopic matter. Given an explicit
expression for $\cL_{k}$, we can get a set of modified Einstein field equations
with Lorentz violation by taking the variation with respect to the metric $g_\mn$.
Though the modified Einstein field equations depend on the specific dynamics of
the Lorentz-violation fields in $\cL_{k}$, it is shown in Ref.\
\cite{Bailey:2006fd}, that in the weak-field regime, the linearized modified
Einstein field equations at the leading order of Lorentz violation can be
expressed using the vacuum expectation values of the Lorentz-violation fields under several reasonable assumptions.
Those vacuum expectation values of the Lorentz-violation fields, denoted as $\bar u$, $\bar s^{\mu\nu}$, and $\bar t^{\alpha\beta\gamma\delta}$, and to be
distinguished from the fields themselves, are called the Lorentz-violation
coefficients. Introducing the Lorentz-violation coefficients makes the SME
framework practically useful by allowing experiments to test the coefficients
without worrying about the dynamics of the corresponding fields as long as gravity
is weak~\cite{Damour:1993hw, Shao:2017gwu, Zhao:2019suc}.
A large amount of constraints have been put on the Lorentz-violation
coefficients from various terrestrial experiments \cite{Chung:2009rm,
Flowers:2016ctv, Shao:2016cjk, Shao:2018lsx} and astrophysical observations
\cite{Shao:2014oha, Shao:2014bfa, Kostelecky:2015dpa, Kostelecky:2016kfm,
Poncin-Lafitte:2016nqd, Yunes:2016jcc, Monitor:2017mdv, Bourgoin:2017fpo,
Shao:2018vul, Shao:2019cyt} that assume the validity of weak gravity.

When applied to neutron stars, the nonlinearity in the Einstein field equations cannot be neglected. Directly using the linearized result in Ref.\ \cite{Bailey:2006fd} leads to a set of Lorentz-violation hydrostatic
equations only at the Newtonian level, causing inaccuracy in describing the structure of neutron stars. Inspired by the post-Tolman-Oppenheimer-Volkoff (post-TOV) approach~\cite{Glampedakis:2015sua, Glampedakis:2016pes}, we mend the inaccuracy by replacing the Lorentz-invariant Newtonian terms with the GR terms in the TOV equation. In this way, the Lorentz-violation terms, once isolated, can be used to calculate the leading-order modification to the structure of neutron stars.

We derive in Sec.\ \ref{mTOVeq} the Lorentz-violation hydrostatic
equations by employing the linearized result in Ref.\ \cite{Bailey:2006fd}. In Sec.\ \ref{perturbationsol}, perturbative expansions of the fluid variables respect to the Lorentz-violation coefficient are performed so that the Lorentz-invariant terms can be identified and replaced by the GR terms in the TOV equation, leaving the Lorentz-violation terms forming the first-order equations. The first-order solution is provided and the calculation to determine the angular dependence is shown in \ref{app}. Finally in Sec.\ \ref{quadrupole}, a spinning neutron star is assumed to estimate the amplitude of the quadrupole radiation from the Lorentz-violation induced deformation. The possibility of using continuous gravitational waves to constrain Lorentz violation is discussed. Throughout the work, we follow the notation and conventions of Ref.\ \cite{Kostelecky:2003fs}.

\section{The Lorentz-violation hydrostatic equations}
\label{mTOVeq}

The TOV equation describes the distribution of the density and pressure inside a perfect
fluid whose local energy-momentum tensor can be expressed as
\bea
T^\mn = (\ep + p) u^\mu u^\nu + p g^\mn \, ,
\label{perfectfluidT}
\eea 
where $\ep$ is the proper energy density and $p$ is the proper pressure of the
fluid. The 4-velocity $u^\mu$ can be taken as $\left(\frac{1}{\sqrt{-g_{00}}},\vec{0} \right)$ for
a static configuration, and then the energy-momentum conservation equations
$D_\mu T^\mn = 0$ give
\bea
\prt_i p + (\ep + p) \frac{\prt_i g_{00} }{2g_{00}} = 0 \, .
\label{gTOV}
\eea
Equation \rf{gTOV} seems to indicate that static structures of fluids only depend on the metric component $g_{00}$. However, other metric components come into the Einstein field equations in solving $g_{00}$, and the presence of the fluid variables themselves in the field equations complicates the problem. In the case of a
spherical fluid, the Einstein field equations imply \cite{Weinberg:1972kfs}
\bea
\frac{\prt_r g_{00} }{2g_{00}} = \left( \frac{Gm(r)}{r^2} + 4\pi G r p \right) \left( 1-\frac{2Gm(r)}{r} \right)^{-1} \, ,
\eea  
with $\prt_\th g_{00} = \prt_\vp g_{00} = 0$ in the Schwarzschild coordinates. The mass function is defined as $m(r)
\equiv 4 \pi \int_0^r \ep \left(r^{\prime} \right) r^{\prime\,2} \dd r^{\prime}$. Hence, Eq. \rf{gTOV} gives
the standard TOV equation
\bea
 \prt_r p = -(\ep + p) \frac{ Gm(r) + 4\pi G r^3 p }{r \left[ r- 2Gm(r) \right] } \, ,
\label{tov} 
\eea
with the angular equations vanishing as expected.

In the case
of the minimal Lorentz violation, the static Newtonian solution for $g_{00}$
obtained in Ref.\ \cite{Bailey:2006fd} reads
\bea
g_{00} = -1 + 2 U + \bar s^{jk} U^{jk} + O( 1{\rm PN} ) \, ,
\label{mSMEg}
\eea
where $\bar s^{jk}$  with $j, k = 1, 2, 3$ are the vacuum expectation values of the spatial components of the Lorentz-violation field $s^{\al\be}$, namely the
spatial components of the Lorentz-violation coefficient $\bar s^{\al\be}$. The
Newtonian potentials $U$ and $U^{jk}$ are defined as 
\bea
 U &=& G\int \frac{1  }{\left| \vec x - {\vec x}^{\,\prime} \right|} \, \rh({\vec x}^{\,\prime}) \, \dd^3 x^{\prime} \, ,
\nonumber \\
 U^{jk} &=& G\int \frac{ (x^j - x^{\prime\,j})(x^k - x^{\prime\,k}) }{\left| \vec x - {\vec x}^{\,\prime} \right|^3} \, \rh({\vec x}^{\,\prime}) \, \dd^3 x^{\prime} \, ,
\eea
where $\rh$ is the baryonic rest mass density. The relative difference between
$\rh$ and $\ep$, namely $\Pi = \frac{\ep - \rh}{\rh}$, is the internal energy per unit
baryonic mass. Note that we have ignored the temporal component $\bar s^{00}$
as it merely rescales the gravitational constant $G$ in a static system and
hence does not produce any observable effect. For the same reason, the
Lorentz-violation coefficient $\bar u$ does not appear and a traceless condition
\bea
\et_{ij} \bar s^{ij} = 0 \, ,
\eea  
can be imposed. As for the absence of the Lorentz-violation coefficient $\bar
t^{\, \al\be\ga\de}$, it is proved in Ref.~\cite{Bailey:2006fd}, that all the
terms involving it automatically cancel out, though a plausible physical
explanation for this remains missing at the moment (the so-called $t$ puzzle)
\cite{Bonder:2015maa}.

Equation \rf{mSMEg} represents the Lorentz-violation solution for $g_{00}$
generated by the couplings in Eq. \rf{mSME} at the Newtonian level. When plugged
into Eq. \rf{gTOV}, we get a set of Newtonian hydrostatic equations with
modifications from the minimal Lorentz violation. These equations are
\bea
 \prt_{r} p &=& \rho \left( \prt_{r} U + \frac{1}{2} \bar s^{jk} \prt_{r} U^{jk} + O(1 {\rm PN}) \right) \, ,
\nonumber \\
 \prt_{\th} p &=& \rho \left( \prt_{\th} U + \frac{1}{2} \bar s^{jk} \prt_{\th} U^{jk} + O(1 {\rm PN}) \right) \, ,
\nonumber \\
 \prt_{\vp} p &=& \rho \left( \prt_{\vp} U + \frac{1}{2} \bar s^{jk} \prt_{\vp} U^{jk} + O(1 {\rm PN}) \right) \, .
\label{newtoniantov}
\eea
The terms in ``$O(1 {\rm PN})$'' include both Lorentz-invariant and Lorentz-violation post-Newtonian corrections. The former are accounted once the TOV equation is used to replace the Lorentz-invariant terms in Eqs. \rf{newtoniantov}, while the latter are neglected as higher-order corrections. Note that the higher-order Lorentz-violation corrections include nonlinear couplings between the Lorentz-violation coefficient and the gravitational field, and might become dominant in certain scenarios where the Lorentz-violation coefficient is fine-tuned so that the couplings blow up in the presence of strong gravitational fields. Our work is restricted to the case where the leading-order Lorentz-violation correction dominates. In the next section, we will identify the Lorentz-invariant terms in Eqs. \rf{newtoniantov} and extract the leading-order Lorentz-violation equations that describe the modification to the structure of the otherwise spherically static perfect fluid.

\section{The leading-order modification to the TOV \mbox{equation} }
\label{perturbationsol}

The fact that any Lorentz-violation effect must be tiny to be consistent with
the experimental support for Lorentz invariance naturally suggests us to treat
the Lorentz-violation terms in Eqs. \rf{newtoniantov} as perturbations. To start, we write the fluid variables $\rh$, $p$, and the potentials $U$ and $U^{jk}$ as perturbation series
\bea
\rh &=& \rh^{(0)} (r) + \rh^{(1)} \left( \vec x \right) + \ldots \, ,
\nonumber \\
p &=& p^{(0)}(r) + p^{(1)}\left( \vec x \right) + \ldots \, ,
\nonumber \\
U &=& U^{(0)} (r) + U^{(1)} \left( \vec x \right) + \ldots \, ,
\nonumber \\
U^{jk} &=& U^{jk \, (0)} \left( \vec x \right) + U^{jk \, (1)} \left( \vec x \right) + \ldots \,  .
\eea 
The zeroth-order fluid variables, $\rh^{(0)} (r)$ and $p^{(0)}(r)$, defined as the solution to Eqs. \rf{newtoniantov} in the absence of Lorentz violation given a proper equation of state
(EOS), satisfy the usual Newtonian hydrostatic equation. 
The first-order corrections, $\rh^{(1)} (r)$ and $p^{(1)}(r)$, are then determined by
\bea
 \prt_{r} p^{(1)} &=& \rh^{(0)} \left( \prt_{r} U^{(1)} + \frac{1}{2} \bar s^{jk} \prt_{r} U^{jk \, {(0)}}  \right)  + \rh^{(1)} \prt_{r} U^{(0)} \, ,
\nonumber \\
 \prt_{\th} p^{(1)} &=& \rh^{(0)} \left( \prt_{\th} U^{(1)} + \frac{1}{2} \bar s^{jk} \prt_{\th} U^{jk \, {(0)}} \right) \, ,
\nonumber \\
 \prt_{\vp} p^{(1)} &=& \rh^{(0)} \left( \prt_{\vp} U^{(1)} + \frac{1}{2} \bar s^{jk} \prt_{\vp} U^{jk \, {(0)}} \right) \, .
\label{1stsmetov}
\eea

Now that we have extracted the equations at the leading-order of Lorentz violation, as we advertised, to account for the relativistic corrections, the usual Newtonian hydrostatic equation at the zeroth order needs to be replaced by the TOV equation \rf{tov}. Namely, $\rh^{(0)} (r)$ and $p^{(0)}(r)$ are now regarded as the solution to the TOV equation, and Eqs. \rf{1stsmetov} describe the leading-order modification to the TOV equation due to Lorentz violation. 

To solve $\rh^{(1)}$ and $p^{(1)}$ in Eqs. \rf{1stsmetov}, let us keep in mind that Lorentz violation not only raises corrections to the
fluid variables but also changes the shape of the fluid, hence the boundary
conditions. Assuming the radius of the fluid sphere $S$ to be $R$ for a given GR
solution, then taking the perturbative change due to Lorentz violation into
consideration, the shape of the fluid can be written as 
\bea
r = \left( 1 + \al(\th, \vp) \right) R \, ,
\label{shape}
\eea
where $\al(\th, \vp)$ is to be determined up to the first order of $\bar
s^{jk}$. Therefore, the boundary conditions for $\rh$ and $p$ are 
\bea
0 &=& \rh \Big|_\Si = \rh^{(0)}\left( R + \al(\th,\vp) R \right) + \rh^{(1)} \left(\vec R \right) \, , 
\nonumber \\
0 &=& p \Big|_\Si = p^{(0)}\left( R + \al(\th,\vp) R \right) + p^{(1)} \left(\vec R \right) \, , 
\label{bc}
\eea
where $\Si$ is the surface described by Eq. \rf{shape} and $\vec R$
represents the position vectors for points on the sphere $S$. 

One difficulty to solve Eqs. \rf{1stsmetov} comes
from the fact that  $U^{(1)}$ depends
nontrivially on $\rh^{(1)}$. We can handle this by writing $U^{(1)}$ as
\bea
U^{(1)} \left( \vec x \right) &=& G\int\limits_{\Si-S} \frac{1}{\left| \vec x - {\vec x}^{\,\prime} \right|} \, \rh^{(0)}(r) \dd^3 x^{\prime} 
\nonumber \\
&& + G\int\limits_{S} \frac{1 }{\left|\vec x - {\vec x}^{\,\prime} \right|} \, \rh^{(1)}({\vec x}^{\,\prime}) \dd^3 x^{\prime} \, .
\label{u1}
\eea
At the first order of $\bar s^{jk}$, the first integral vanishes because it can
be approximated at $r=R$ and $\rh^{(0)}(R) = 0$ is guaranteed in GR solutions.
Another difficulty is that since $\rh^{(1)}$ and $p^{(1)}$ are Lorentz-violation induced anisotropic
corrections, the usual isotropic EOS that relates $p^{(0)}$ and $\rh^{(0)}$ does
not apply to them. This issue is compensated by the requirement that the second derivatives of $p^{(1)}$
exist, which implies 
\bea
\hspace{-0.4cm}
\prt_\th  \rh^{(1)} \prt_{r} U^{(0)} &=&  \prt_r \rh^{(0)}  \left( \prt_{\th} U^{(1)} + \frac{1}{2} \bar s^{jk} \prt_{\th} U^{{(0)}\,jk} \right) \, ,
\nonumber \\
\hspace{-0.4cm}
 \prt_\vp  \rh^{(1)} \prt_{r} U^{(0)} &=&  \prt_r \rh^{(0)} \left( \prt_{\vp} U^{(1)} + \frac{1}{2} \bar s^{jk} \prt_{\vp} U^{{(0)}\,jk} \right) \, .
\label{rhgr1}
\eea

\begin{figure*}
    \centering
    \includegraphics[width=5cm]{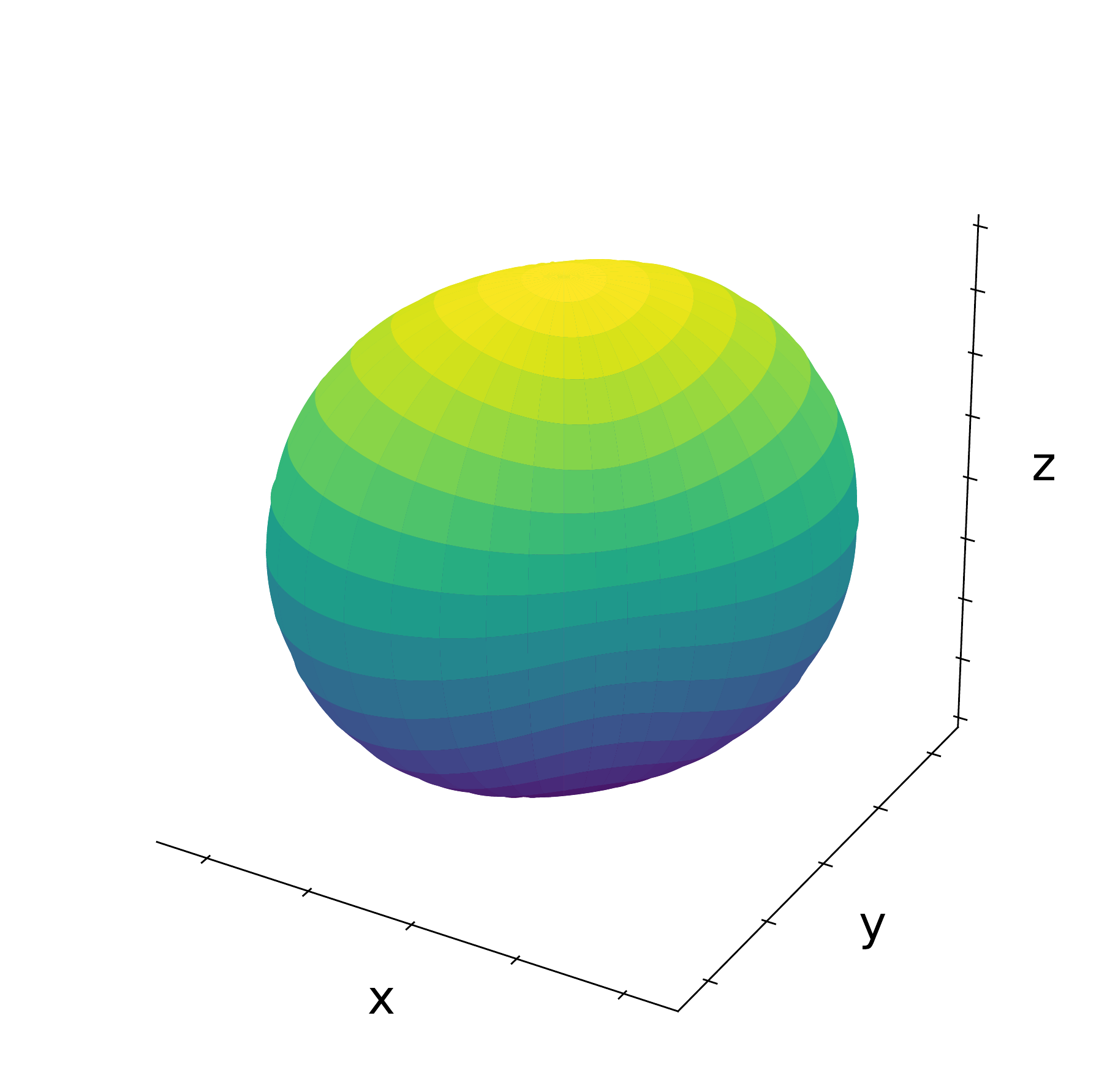}
    \includegraphics[width=5cm]{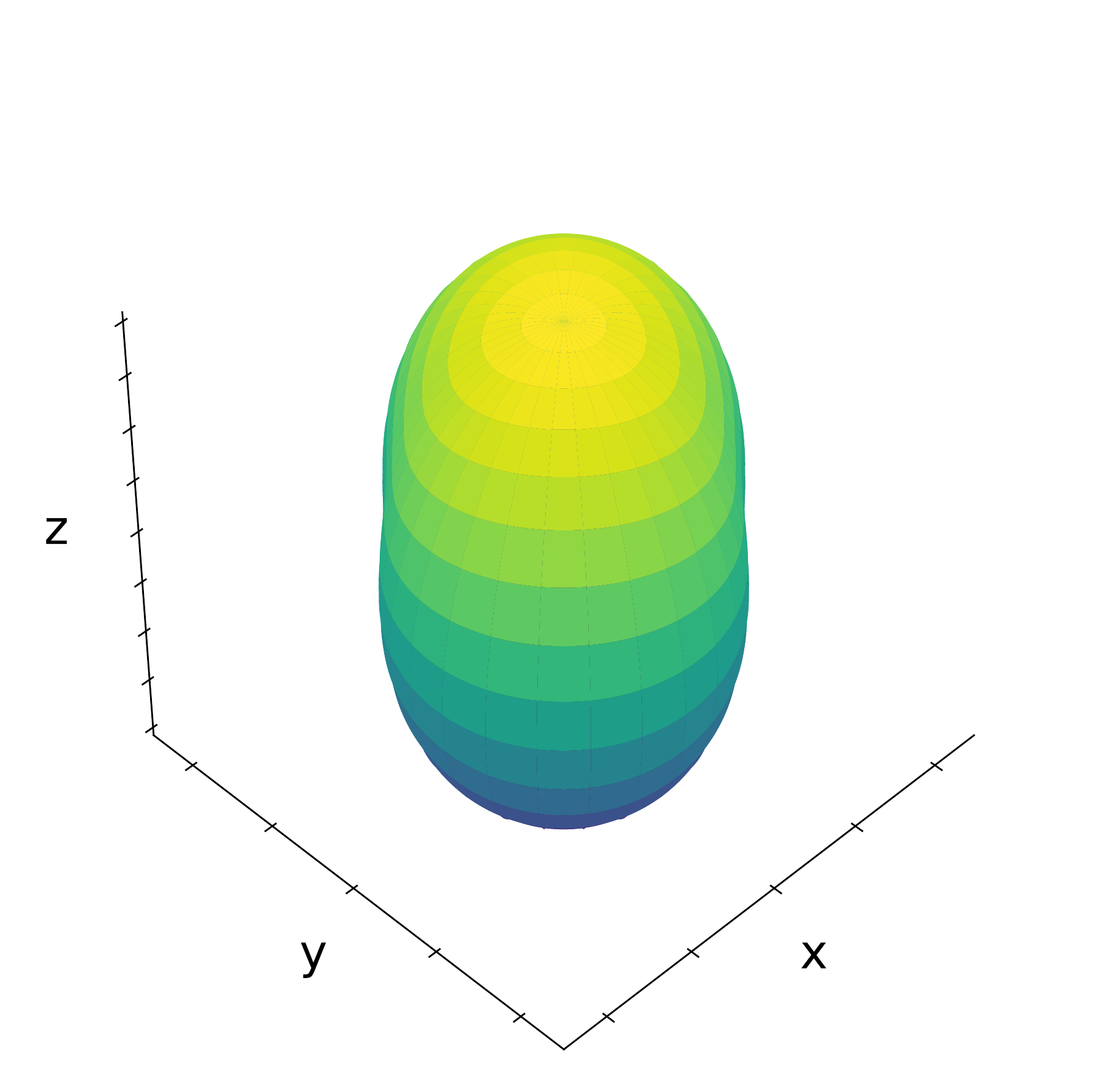}
    \includegraphics[width=5cm]{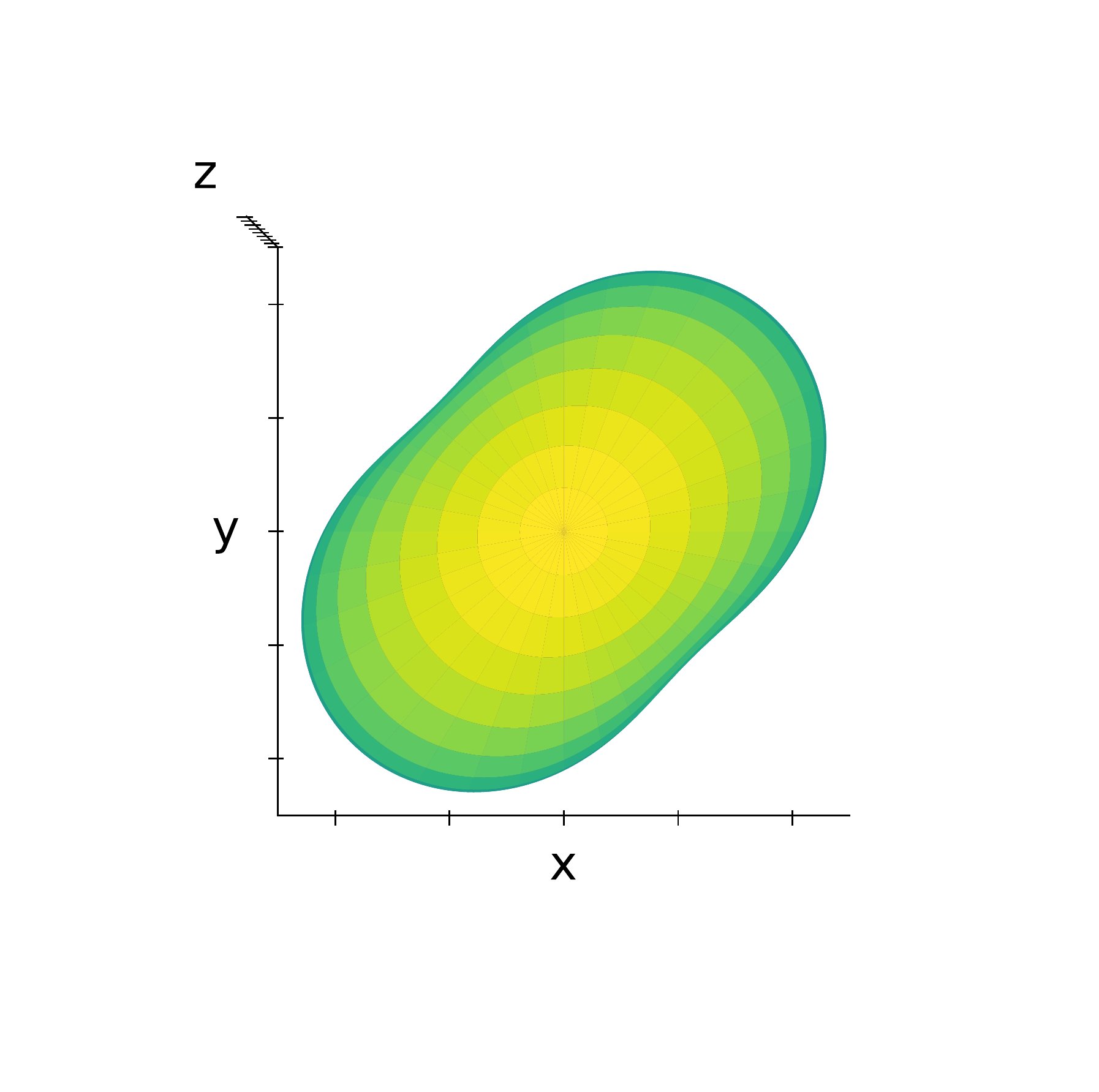}
    \caption{Different views of a neutron star with minimal Lorentz
    violation. For illustrative purpose, we choose $\bar{s}^{xy} = 0.5$ and all the other components of the Lorentz-violation coefficient $\bar s^{jk}$ vanish.}
    \label{fig:my_label}
\end{figure*}

Skipping the tedious calculation using series expansion, we display the surprisingly tidy and simple perturbation solution
\bea
 p^{(1)}\left( \vec x \right) &=& -\al(\th, \vp) r\, \prt_r p^{(0)} (r) \, , 
\nonumber \\
 \rh^{(1)} \left( \vec x \right) &=& -\al(\th, \vp) r\, \prt_r \rh^{(0)} (r) \, .
\label{sol}
\eea
It is straightforward to verify that the solution \rf{sol} satisfies Eqs.
\rf{1stsmetov} and Eqs. \rf{rhgr1}, as long as the spherical harmonic expansion of
$\al(\th, \vp)$ is (see~\ref{app})
\bea
\al(\th, \vp) = \frac{1}{2} \sum\limits_{m= -2}^{2} s^{(s)}_{2m} Y_{2m} (\th, \vp) \, ,
\label{al}
\eea
where $s^{(s)}_{2m}$ are the spherical components of $\bar s^{jk}$. The explicit
relations between $s^{(s)}_{2m}$ and the cartesian components are
\bea
&& s^{(s)}_{2,-2} = \sqrt{ \frac{ 2\pi }{15}} ( \bar s^{xx} - \bar s^{yy} + 2i \bar s^{xy} ) \, ,
\nonumber \\
&& s^{(s)}_{2,-1} = 2 \sqrt{ \frac{2\pi}{15} } ( \bar s^{xz} + i \bar s^{yz} ) \, ,
\nonumber \\
&& s^{(s)}_{2,0} = \frac{2}{3} \sqrt{\frac{\pi}{5} } ( -\bar s^{xx} - \bar s^{yy} + 2 \bar s^{zz} )  \, ,
\nonumber \\
&& s^{(s)}_{2,1} = 2 \sqrt{ \frac{2\pi}{15} } ( - \bar s^{xz} + i \bar s^{yz} ) \, ,  
\nonumber \\
&& s^{(s)}_{2,2} = \sqrt{\frac{ 2\pi }{15}} ( \bar s^{xx} - \bar s^{yy} - 2i \bar s^{xy} ) \, .
\eea
In addition, it is clear that the boundary conditions \rf{bc} are also satisfied
by the solution \rf{sol} given $\rh^{(0)}(R) = p^{(0)}(R) = 0$.

Before applying it to neutron stars, we would like to clarify that the solution
\rf{sol} is physical though its particular form suggests that it can be
generated from the Lorentz-invariant quantities $p^{(0)}(r)$ and $\rh^{(0)}(r)$ by a coordinate transformation
\bea
r \rightarrow r^{\prime} = \left( 1 - \al(\th, \vp) \right) r \, .
\eea
The inverse transformation seems to eliminate the Lorentz-violation corrections,
but actually just hides the effects into the spatial part of the metric. Taking the
Newtonian limit as an example, the boundary of the fluid becomes the sphere $r^{\prime}
= R$ in the $(r^{\prime}, \th, \vp)$ coordinates. However, the spatial part of the metric in these
coordinates are
\bea
g'_{jk} = 
\begin{pmatrix}
1+2\al & r^{\prime} \prt_\th \al  & r^{\prime} \prt_\vp \al\\
r^{\prime} \prt_\th \al & r^{\prime 2} \left( 1 + 2 \al \right)  & 0\\
r^{\prime} \prt_\vp \al & 0 & r^{\prime 2} \left( 1 + 2 \al \right) \sin^2\th 
\end{pmatrix} \, ,
\eea  
with $\al(\th, \vp)$ describing the same Lorentz-violation effects as one would
experience in the coordinates $(r, \th, \vp)$ where $g_{jk} = \eta_{jk}$. 

\begin{figure}
    \centering
    \includegraphics[width=6cm]{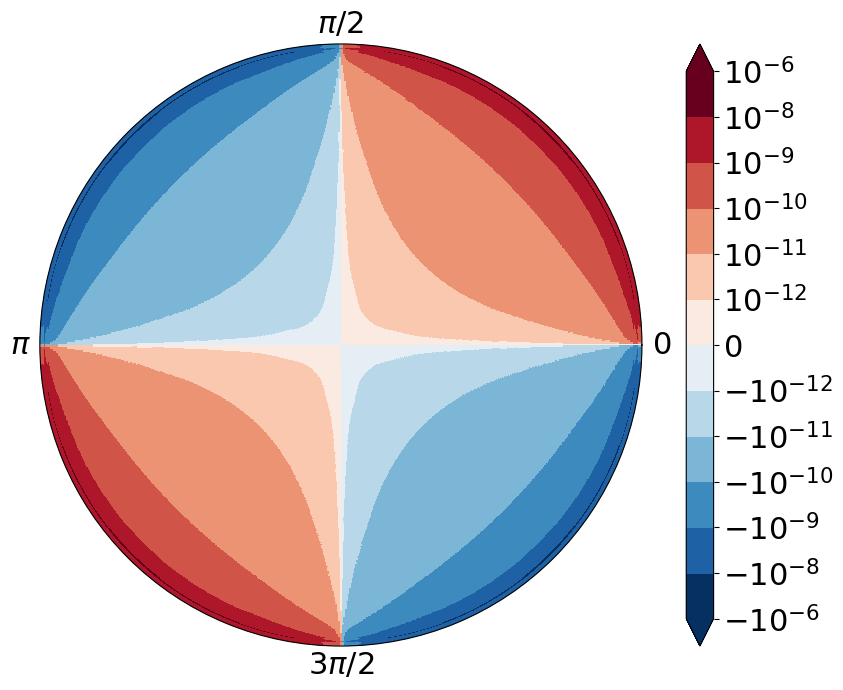}
\caption{The distribution of the relative density correction, ${\rho^{(1)}(\vec
x)}/{\rho^{(0)}(r)}$, in the equatorial section (the $X$-$Y$ plane) of a neutron star. We have
assumed that all the other independent components of the Lorentz-violation
coefficient vanish except for $\bar{s}^{xy} = 10^{-10}$. The zeroth-order
solution is obtained numerically with the EOS {\sf AP4} for a neutron star with mass $1.44\, {M_{\odot}}$.}
    \label{fig:Sxy_XY}
\end{figure}

\section{Newtonian quadrupole of a neutron star}
\label{quadrupole}

Deformed neutron stars emit continuous gravitational waves when rotating
\cite{Poisson:2014misc}. The quadrupole radiation is the leading term in the
post-Newtonian expansion. As the quadrupole moments themselves are defined at
the Newtonian level using the baryonic rest mass density $\rh$, the solution
\rf{sol} is just accurate to give us the quadrupole moments caused by the
minimal Lorentz violation. 

To illustrate the effect of Lorentz violation on the deformation of a neutron star, we plot in Fig.~\ref{fig:my_label} the shape viewed from three different angles with an unrealistically large component of the Lorentz-violation coefficient $\bar s^{jk}$. In addition, in Fig.~\ref{fig:Sxy_XY} we plot the fractional correction to the density of a neutron star, whose mass is fixed to $1.44\, {M_{\odot}}$ with the EOS {\sf AP4}~\cite{Lattimer:2000nx}. The more relevant value chosen for the component of $\bar s^{jk}$ in Fig.~\ref{fig:Sxy_XY} is based on its current bounds in Ref.\ \cite{Kostelecky:2008ts}. 

Figures \ref{fig:my_label} and \ref{fig:Sxy_XY} show that in general both the shape of
the star and the density of the star become anisotropic under the influence of
Lorentz violation. This indicates anisotropic quadrupole moments. Using the solution
\rf{sol}, the quadrupole moments are found to be
\bea
\hskip -0.5cm I^{jk} &=& \int\limits_\Si x^j x^k \left( \rh^{(0)}(r) + \rh^{(1)}\left( \vec x \right) \right) \dd^3 x 
\nonumber \\
&=& \frac{1}{3}  \left( \de^{jk} + \bar s^{jk} \right) I \, ,
\eea
where $\de^{jk}$ is the Kronecker delta and $ I = 4\pi \int_0^R r^4 \rh^{(0)} \dd r$ is the rotationally invariant trace of the quadrupole moments. The anisotropy related
to the quadrupole radiation is measured by the ellipticity
\bea
e = \frac{ I^{YY} - I^{XX} }{I^{XX} + I^{YY} } \, ,
\label{ellip}
\eea
where $ \{ I^{XX} , \, I^{YY} , \, I^{ZZ}\} $ are the eigenvalues of $I^{jk}$, and $I^{jk}$ is diagonalized in the $\left( X,\, Y,\, Z \right)$ coordinates. The ellipticity \rf{ellip} applies to rotations along the $Z$-axis, and the case for a general spin direction can be obtained with 3-dimensional rotations.

As an example, we consider the quadrupole radiation of a deformed neutron star due to $\bar s^{xy}$ alone spinning in the $z$-direction. The coordinates that diagonalize $I^{jk}$ have the $Z$-axis along the $z$-axis, while the $X$ and $Y$ coordinates are related to $(x,\,y)$ by the coordinate transformation
\bea
X &=& \frac{1}{\sqrt{2} } \left( x + y \right) ,
\nonumber \\
Y &=& \frac{1}{\sqrt{2} } \left( x - y \right) .
\eea
The eigenvalues of $I^{jk}$ are
\bea
I^{XX} &=& \frac{1}{3} \left( 1 + \bar s^{xy} \right) I ,
\nonumber \\
I^{YY} &=& \frac{1}{3} \left( 1 - \bar s^{xy} \right) I ,
\nonumber \\
I^{ZZ} &=& I^{zz} = \frac{1}{3} I .
\eea
Then, the ellipticity associated with rotations along the $Z$-axis is simply $\bar s^{xy}$ at the leading order of Lorentz violation, and the amplitude
of the quadrupole radiation can be estimated as \cite{Poisson:2014misc}
\bea
h_0 &=&  \frac{4G \Om^2}{d} \frac{2}{3} I \bar s^{xy} 
\nonumber \\
&\simeq& 7 \times 10^{-28}  \left( \frac{1 {\rm ms} }{P} \right)^2  \left(
\frac{1 {\rm kpc} }{d} \right) \left(\frac{\bar s^{xy}}{10^{-10}}\right)  ,
\eea
where $\Om = \frac{2\pi}{P}$ is the angular velocity of the neutron star with $P$ being the spin period, and $d$ is the distance of the neutron star. To obtain the approximate numerical value, we used $ M = 1.4 \, M_{\odot}$ for the mass of the neutron star and $R=12$~km for its radius. A uniform density is assumed to estimate the trace of the quadrupole moments, namely $I = \frac{3}{5} M R^2 = 2.4 \times 10^{38} \, {\rm kg\, m^2}$. Note that the moment of inertia along any diameter of the uniform sphere is $\frac{2}{3} I$. 

Continuous gravitational waves are important signals for the LIGO/Virgo
detectors. Various algorithms are being developed for possible
events~\cite{ Whelan:2015dha, Ming:2015jla, Walsh:2016hyc, Abbott:2017mwl, Abbott:2017hbu, Dreissigacker:2018afk, 
Sun:2019bew, Authors:2019ztc, Dergachev:2019wqa, Dergachev:2019oyu}. Although there is no detection yet,
meaningful constraints are already set from the advanced detectors for
different systems at levels of $h_0 \lesssim
10^{-25}$~\cite{Abbott:2017mwl, Pisarski:2019vxw, Abbott:2019uwg, Dergachev:2019wqa, Dergachev:2019oyu} and $e
\lesssim 10^{-9}$--$10^{-8}$~\cite{Abbott:2018qee, Authors:2019ztc, Ming:2019xse}. The
estimated quadrupole radiation from Lorentz violation is too weak to be
detected currently. The idea of Lorentz violation, anyway, provides a
possible cause of continuous gravitational waves for future observations.
Also, it is important to keep in mind that current constraints on $\bar
s^{jk}$~\cite{Kostelecky:2008ts} all assumed experiments and observations
involving only weak gravitational fields. We expect Lorentz violation to be
comparatively larger in a strong gravitational scenario. It is possible
that the quadrupole radiation due to Lorentz violation might be comparable
to or even greater than that of conventional deformations, like a mountain
on the star or the tidal interactions in a close binary
\cite{Poisson:2014misc}. Despite the fact that the amplitude of the quadrupole radiation does not distinguish a Lorentz-violation deformation from conventional deformations, upper bounds on Lorentz violation in strong-field
systems will be told once any continuous gravitational
waves are detected in the future.   

Finally, there remains a natural and essential question on whether there are any signatures in the continuous gravitational
waves that distinguish Lorentz violation from conventional deformations when more stringent constraints on Lorentz violation are to be extracted. The answer requires a full Lorentz-violation quadrupole radiation formula which in principle follows the 1PN metric solution with Lorentz violation and a nonstatic fluid configuration. Therefore, any Lorentz-violation signature in the quadrupole radiation is a second-order effect in the sense that the Lorentz-violation coefficient couples with 1PN terms instead of the Newtonian term as we study here. Such effects most likely show up in the phases and polarizations of the waves as the amplitude is dominated by the first-order effect that we have calculated. While a detailed derivation is beyond the scope of the present work, the question is certainly worth an investigation.

\section{Summary}
\label{sum}

We calculated the leading-order modification to a static perfect fluid due to the minimal Lorentz violation and applied it to neutron stars to find an estimate for the quadrupole radiation. The corrections were first derived at the
Newtonian level as shown in Eqs. \rf{newtoniantov}. Then, the Lorentz-violation corrections and the Lorentz-invariant terms are separated using the perturbation method, and the TOV equation replaces the zeroth-order Lorentz-invariant hydrostatic equation to account for relativistic effects in strong gravitational systems like neutron stars. The perturbation solution \rf{sol} to the first-order equations \rf{1stsmetov} was found and used to estimate the quadrupole gravitational radiation for continuous gravitational waves. Our
calculation shows that the amplitude is too weak to be detectable at the moment,
but we expect that future observations of continuous gravitational waves can
make use of our result and set constraints on Lorentz violation in the
strong-field regime.

We point out that it is possible, if not straightforward, to generalize
our results to the nonminimal gravitational SME~\cite{Bailey:2014bta}. We conjecture that the solution still takes the form of
\rf{sol} with higher spherical harmonics involved in the angular function $\al$. The specific dependence of the factors in front of the spherical
harmonics on the nonminimal Lorentz-violation coefficients as an analog of Eq. \rf{al} requires further calculation.

\section*{Acknowledgements}
We are grateful to Quentin G. Bailey and M. Alessandra Papa for comments. This work was supported by the National Natural Science Foundation of China (11975027), the Young Elite Scientists Sponsorship Program
by the China Association for Science and Technology (2018QNRC001), and the
High-performance Computing Platform of Peking University. It was partially
supported by the National Natural Science Foundation of China (11721303),
and the Strategic Priority Research Program of the Chinese Academy of
Sciences through the Grant No. XDB23010200. R.X. is supported by the Boya
Postdoctoral Fellowship at Peking University.

\appendix
\section{Determining $\al(\th, \vp)$}
\label{app}

Substituting the solution \rf{sol} into Eqs. \rf{1stsmetov} and Eqs.
\rf{rhgr1}, we find that $\al(\th,\vp)$ only needs to satisfy
\bea
-\al r \prt_r U^{(0)} = U^{(1)} + \frac{1}{2} \bar s^{jk} U^{{(0)}\,jk} \, .
\label{aleq}
\eea
The spherical expansion of the left-hand side is 
\bea
\hskip -0.4cm -\al r \prt_r U^{(0)} &= & 4\pi G\sum\limits_{l,m} \al_{lm} Y_{lm} (\th, \vp)  \frac{1}{r} 
\nonumber \\
&& \int_0^r r^{\prime\, 2} \rh^{(0)} \left(r^{\prime} \right)  \dd r^{\prime} \, ,
\eea
and the spherical expansions of the terms on the right-hand side are
\bea
U^{(1)} &=& 4\pi G \sum\limits_{l,m} \frac{4\pi}{2l+1} \al_{lm} Y_{lm} (\th, \vp)
\nonumber \\
&& \left( \frac{l+3}{r^{l+1}} \int_0^r r^{\prime\, l+2} \rh^{(0)} (r^{\prime}) \dd r^{\prime} \right.
\nonumber \\
&& \left. - (l-2)r^l \int_r^R \frac{\rh^{(0)}(r^{\prime})}{r^{\prime\,l-1} } \dd r^{\prime} \right) \, ,
\eea
and
\bea
\bar s^{jk} U^{{(0)}\,jk} &=&  4\pi G \sum\limits_{m} s^{(s)}_{2m} Y_{2m} (\th,\vp)  
\nonumber \\
&& \left( \frac{1}{r} \int_0^r  r^{\prime\,2}  \rh^{(0)} (r^{\prime})  \dd r^{\prime} \right.
\nonumber \\
&& \left. - \frac{1}{r^3} \int_0^r r^{\prime\,4}  \rh^{(0)} (r^{\prime})  \dd r^{\prime} \right) \, .
\eea
Note that we have used the solution \rf{sol} to compute $U^{(1)}$ from Eq. \rf{u1} and the condition $\et_{jk} \bar s^{jk} = 0 $ to simplify $\bar s^{jk} U^{(0)\,jk}$.  
By comparing, we obtain 
\bea
\al_{lm} = 
\begin{cases}
    \frac{1}{2} s^{(s)}_{2m}, & \text{for } l = 2 \, ,\\
    0, & \text{for } l \ne 2 \, .
\end{cases}
\eea

\bibliography{mybibfile}

\end{document}